\documentclass[b5paper,twoside]{jpconf}  
\usepackage{graphicx}

\begin{document}

\title[Simulations of Galactic Outflows]
{Numerical Simulations of Galactic Outflows and 
Evolution of the IGM}  

\author[Hugo Martel]{Hugo Martel$^{1,2}$}

\address{$^1$D\'epartement de physique, de g\'enie physique et d'optique,
Universit\'e Laval, Pavillon Alexandre-Vachon, Quebec City, QC, G1V 0A6, Canada}
\address{$^2$Centre de Recherche en Astrophysique du Qu\'ebec}

\ead{hmartel@phy.ulaval.ca} 

\begin{abstract}
Galactic outflows play a major role in the evolution of galaxies
and the intergalactic medium (IGM). The energy deposited into the interstellar
medium by supernovae and active galactic nuclei can accelerate the gas past
the escape velocity, and eject it into the IGM. This will affect the
subsequent evolution of the galaxy, by reducing or eliminating star formation,
and quenching the accretion of matter onto the central AGN. Galactic outflows 
is the main process by which energy and processed interstellar matter is
transported into the IGM. This affects the subsequent formation of other
galaxies. The energy carried by outflows can strip protogalactic halos of their
gas, preventing galaxies from forming. Conversely, the metals carried by
outflows can modify the composition and cooling rates of the gas in
protogalactic halos,
favoring the formation of galaxies. In this paper, I review the various
techniques used to simulate galactic outflows and their impact
on galaxy and IGM evolution.
\end{abstract}

\section{Galactic Outflows}
Galactic outflows are the primary mechanism by which galaxies
deposit energy and metal-enriched gas into the
IGM. This can greatly affect the evolution of the IGM, and the subsequent
formation of other generations of galaxies. 
Feedback by galactic outflows can provide an explanation for
the observed high mass-to-light ratio of dwarf galaxies and the
abundance of dwarf galaxies in the Local Group,
and can solve various problems with galaxy formation models, such as
the overcooling problem and the angular momentum problem
(see \cite{benson10} and references therein).
Galactic outflows can explain
the high metal content ($0.1-1\% Z_\odot$) of the IGM, observed
via the Lyman-$\alpha$ forest
\cite{my87,schayeetal03,ph04,aguirreetal08,pierietal10a,pierietal10b},
the high metal content ($0.3Z_\odot$) of the intracluster medium
in massive X-ray clusters, and the high entropy
content and scaling relations in X-ray clusters
\cite{kaiser91,eh91,cmt97,tn01,babuletal02,voitetal02}.
They also provide observational tests that can constrain 
theoretical models of galaxy evolution. Local examples of spectacular 
outflows in dwarf starburst galaxies include those of the extremely 
metal-poor I Zw 18 \cite{pequi08,jamet10} and NGC1569 \cite{west09}.
More massive spirals, such as NGC7213 \cite{hameed01}, also show 
evidence of global outflows. Figure 1 shows a composite image of the
galaxy M82. The edge-on disc of the galaxy appears in the optical, while 
the gas expelled by the galaxy is seen in $\rm H\alpha$. That gas
forms a bipolar outflow aligned along the direction normal to the
plane of the galaxy (image courtesy Smith, Gallagher, \& Westmoquette).

\begin{figure}
\begin{center}
\includegraphics[]{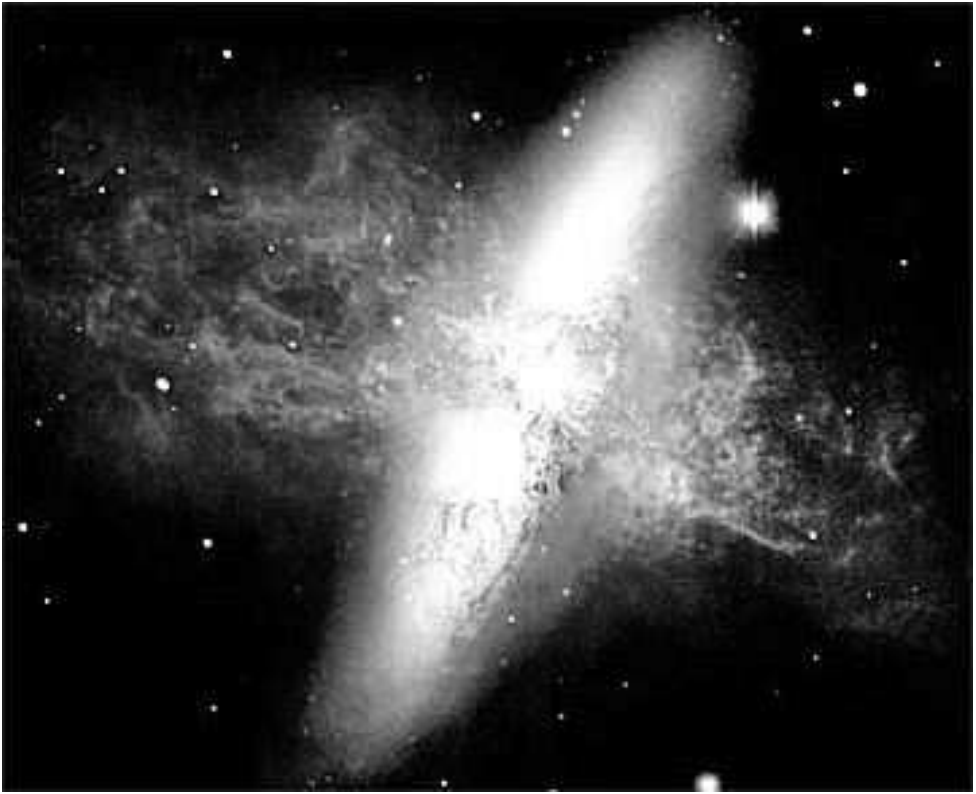}
\end{center}
\caption{\label{M82}Composite image of M82, showing the edge-on galactic 
disk and the bipolar galactic outflow.}
\end{figure}

\section{Joint Evolution of Galaxies and the IGM}
\begin{figure}
\begin{center}
\includegraphics[width=3.7in,angle=270]{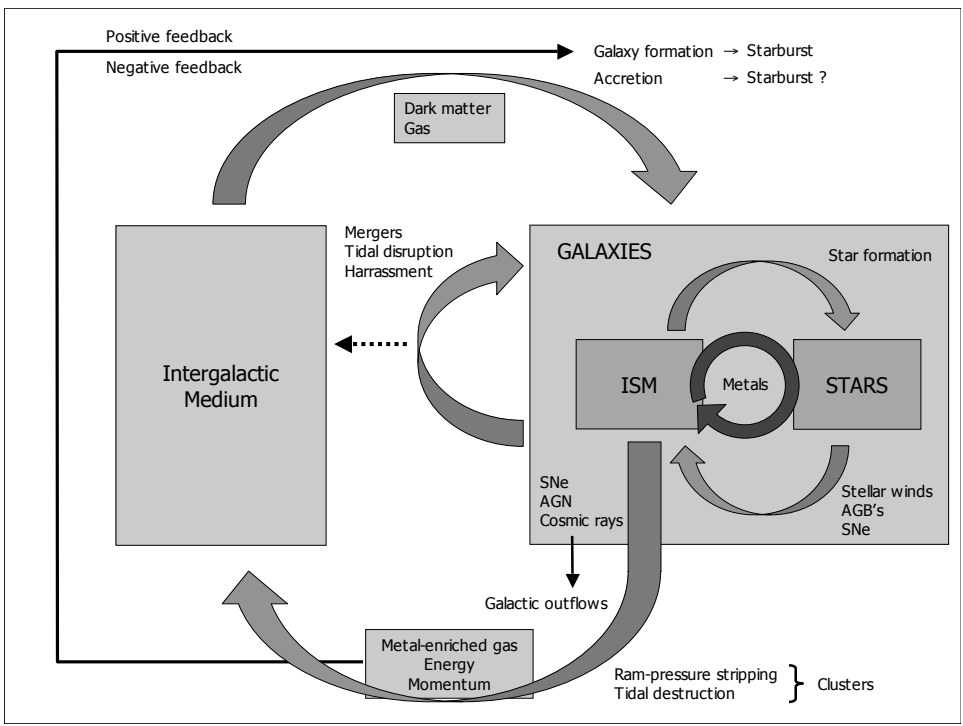}
\end{center}
\caption{\label{bigfigure}Summary of the various physical
processes driving the evolution of galaxies and the IGM.}
\end{figure}

Figure~\ref{bigfigure} summarizes the various physical processes
driving the evolution of galaxies and the IGM. Gas and dark matter in the IGM
is converted into galaxies by the galaxy formation process. This normally 
results in a starburst: the rapid formation of a large number of stars in the 
newly-formed galaxy. Later-on, additional IGM matter can be accreted onto 
existing galaxies, possibly resulting in additional starbursts. Within galaxies,
star formation converts ISM gas to stars, and some of that gas is eventually
returned to the ISM by stellar winds, AGBs and SNe, resulting in a
metallicity increase both in the ISM and in stars. When the energy deposited
into the ISM by SNe, AGNs, and cosmic rays is sufficiently large, energy,
momentum, and metal-enriched gas can be ejected from the galaxy and deposited 
into the surrounding IGM. This can have either a positive or a negative
feedback effect of the subsequent formation of other galaxies. Note that
galactic outflows are not the only process by with galaxies can return matter
into the IGM. Interactions between galaxies might result in mergers, tidal 
disruption, or harassment, in which cases some of the content of galaxies
(both gas and stars) can
be dispersed into the IGM. Also, ram-pressure stripping and tidal destruction
can have the same effect, though they normally do not involve field
galaxies, and only take place in massive clusters.

\section{Individual Outflows: Analytical Models}
The propagation of galactic outflows into the IGM can be described
with a simple analytical model\cite{tse93,sfm02}.
In this model, the injection of thermal energy into the ISM
produces an outflow of radius $R$, which consists of a dense shell
of thickness $R\delta$ containing a cavity.
A fraction $1-f_m$ of the IGM gas initially located
inside radius $R$ is piled up in the shell, while a
fraction $f_m$ of that gas is distributed inside the cavity.
We normally assume $\delta\ll1$, $f_m\ll1$, that is, 
most of the gas is located inside a thin shell.
This is called the {\it thin-shell approximation}.

The evolution of the shell radius $R$ expanding out of a galaxy
of mass $M_{\rm gal}$, is described
by the following system of equations:
\begin{eqnarray}
\label{rdotdot}
\ddot R&=&{8\pi G(p-p_{\rm ext})\over\Omega_bH^2R}-{3\over R}(\dot R-HR)^2
-{\Omega H^2R\over2}-{GM_{\rm gal}\over R^2}\,,\\
\label{pdot}
\dot p&=&{L\over2\pi R^3}-{5\dot Rp\over R}\,,
\end{eqnarray}

\noindent where a dot represents a time derivative, $\Omega$, $\Omega_b$,
and $H$ are the total density parameter, baryon density parameter, and
Hubble parameter at time $t$, respectively,
$L$ is the luminosity (rate of energy injection), 
$p$ is the pressure inside the cavity
resulting from this luminosity, and $p_{\rm ext}$ is the external
pressure of the IGM. 
The four terms in equation~(\ref{rdotdot}) represent, from left to right, the 
driving pressure of the outflow, the drag due to sweeping up the IGM and 
accelerating it from velocity $HR$ to velocity $\dot R$, and the gravitational
deceleration caused by the expanding shell and by the halo itself.
The two terms in equation~(\ref{pdot}) represent the increase
in pressure caused by injection of thermal energy, and the drop in 
pressure caused by the expansion of the shell, respectively. 
To use this model, we need to specify the sources of energy
(SNe, AGNs, cosmic rays,...) driving the outflow, in order
to determine the rate of energy injection $L$. We also need to
specify the external pressure $p_{\rm ext}$ of the IGM into which
the outflow is propagating, which depends upon various approximations
about the composition, temperature, and ionization state of the
IGM\cite{sb01,pmg07}.

\section{Individual Outflows: Simulations}

Numerical simulations of outflows from
isolated dwarf galaxies have been performed by
MacLow and Ferrara\cite{mf99}. In these simulations, they
start with a disc galaxy in equilibrium. They then 
add a continuous source 
of energy in the center of the galaxy (hence this is not a starburst).
The energy added drives the expansion of a bipolar outflow that propagates
in the direction normal to the disk of the galaxy.
For some combinations of the parameters, they found 
that the entire ISM is ejected and escapes the system.

One limitation of these simulations is that galaxies are treated as isolated.
Martel \& Shapiro\cite{ms01} simulated the formation of a galactic halo
at the intersection of two filaments inside a cosmological pancake. The halo
grows by gravitational instability, and when the gas density reaches a certain
threshold, a large amount of thermal energy is 
deposited into the gas, simulating a starburst.
That energy drives the expansion of an outflow.
The galaxy is isolated (that is, not surrounded by other galaxies), but
the cosmological environment is taken into account. The galaxy can accrete
matter from the surrounding structures, and these structures influence the
propagation of the outflow. These simulations show that 
galactic outflows tend to be anisotropic, and propagate along the direction
of least resistance. In some cases, the entire ISM is blown away by the
outflow, as in \cite{mf99}, but accretion for the filaments can rebuild
the gas content of the galaxies, possibly leading to additional starbursts.

\section{Cosmological Simulations}
Simulations of isolated galaxies have several limitations.
They do not describe the effect of the outflow on the
evolution of the IGM and the subsequent formation of other galaxies.
They do not account (usually) for the effect of surrounding structures 
(filament, pancakes, other galaxies) on the propagation of the outflow.
Finally, they do not describe the global effect of the galaxy population on 
the evolution  of the IGM (filling fraction of metals, distribution 
of metals in the IGM, cross-pollution between galaxies, ...).
To address these issues, we need to perform full cosmological simulations, 
with an entire galaxy population, inside a cosmological volume representative
of the universe. This causes a major problem 
with dynamical range: cosmological simulations simply
cannot resolve the galactic
scales, where the physical processes generating the outflows are
taking place. 

The solution is to combine numerical simulations 
of large-scale structure with a 
subgrid treatment of galaxy formation and outflows.
A numerical algorithm is used to simulate large-scale structure formation.
During the simulation, the algorithm
determines where and when stars and/or galaxies form, using some
criterion. In gravity-only simulations, galaxies form when the matter density
$\rho_{\rm DM}^{\phantom1}$ exceeds a certain threshold $\rho_{\rm thres}$. In 
hydrodynamical simulations, galaxies form in regions where the gas density
is high and the temperature is low, $\rho_{\rm gas}>\rho_{\rm thres}$ 
and $T<T_{\rm thres}$.

Once the galaxies have been located in the simulation, the algorithm can 
simulate the propagation of outflows and their effect on the nearby IGM matter.
There are two possible approaches for doing this: 
analytical, or subgrid physics.

\subsection{Numerical $+$ Analytical}

\begin{figure}
\begin{center}
\includegraphics[width=5.4in]{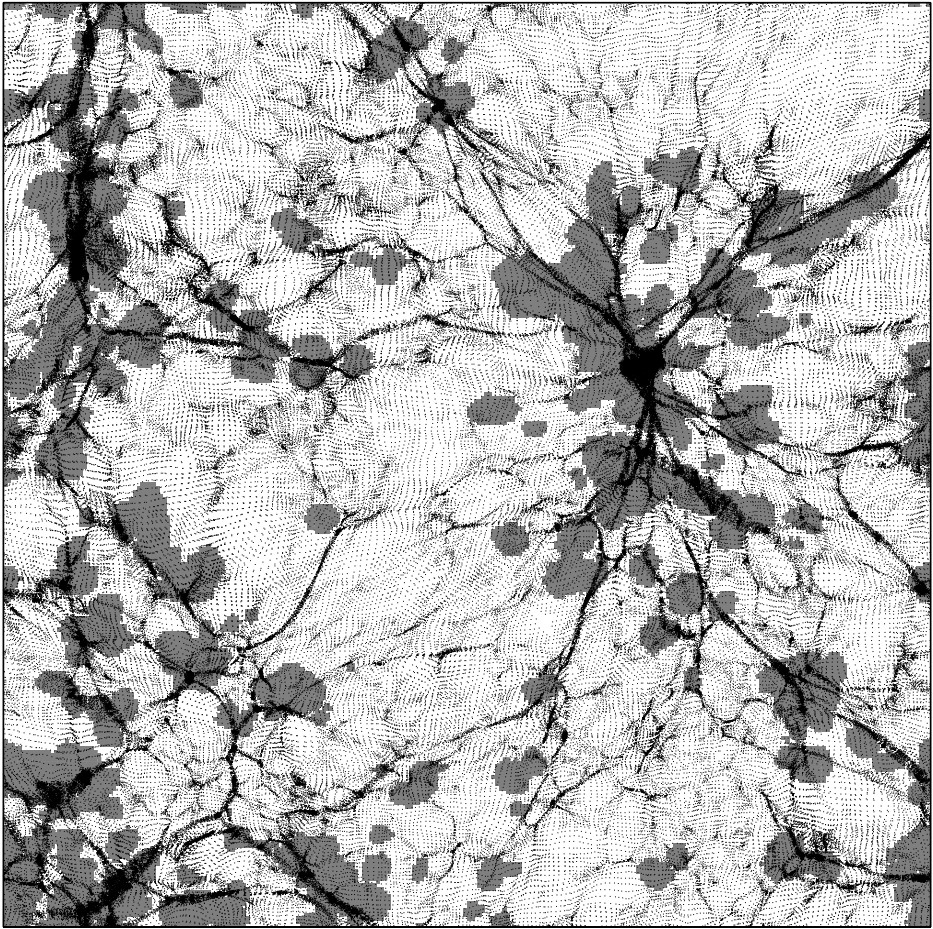}
\end{center}
\caption{Cosmological simulation of structure formation and outflows
in a $\Lambda$CDM universe, inside a computational volume $\rm(15\,Mpc)^3$.
The simulation combines a numerical
algorithm for the formation of large-scale structures with an analytical
model for outflows.
The figure shows the matter distribution (black particles)
and the metal-enriched gas deposited in the IGM by outflows (gray areas),
inside a slice of comoving thickness $\rm0.1\,Mpc$, at redshift $z=2$. 
\label{slice}
}
\end{figure}

The analytical approach consists of ``painting'' an
analytical solution for outflows on a numerical solution of
galaxy and large-scale structure formation. Once a galaxy 
forms, the algorithm calculates 
analytically the propagation of the outflow originating 
from that galaxy, taking into account the mass of the galaxy, its formation
redshift, and the surrounding IGM. Early simulations\cite{sb01,tms01,bsw05}
used the basic Tegmark et al. analytical model\cite{tse93},
which assumes that the outflow propagates into a uniform IGM.
This model was later modified to account for the radial\cite{lg05}
or angular\cite{pmg07,gbm09,pmp10}
variations of the 
IGM density and pressure. Figure~\ref{slice} shows a simulation
of IGM enrichment, which combines an N-body simulation with
an analytical outflow model\cite{pmp10}. The black dots represent the particles,
and show a network of clusters connected by filaments. 
These dense regions are the sites of galaxy formation. The gray areas represent
the metal-enriched gas that has been deposited into the IGM by outflows.

These algorithms take into account, not only the enrichment of the IGM,
but also the effect of outflows on other galaxies.
If a protogalaxy is hit by an outflow, the algorithm estimates whether
ram-pressure stripping is sufficient to prevent the galaxy from forming.
If it is not, then the protogalaxy is enriched in metals, which reduces
the cooling rate of the gas.

\subsection{Numerical $+$ Subgrid Physics}

\begin{figure}
\begin{center}
\includegraphics[width=5.8in]{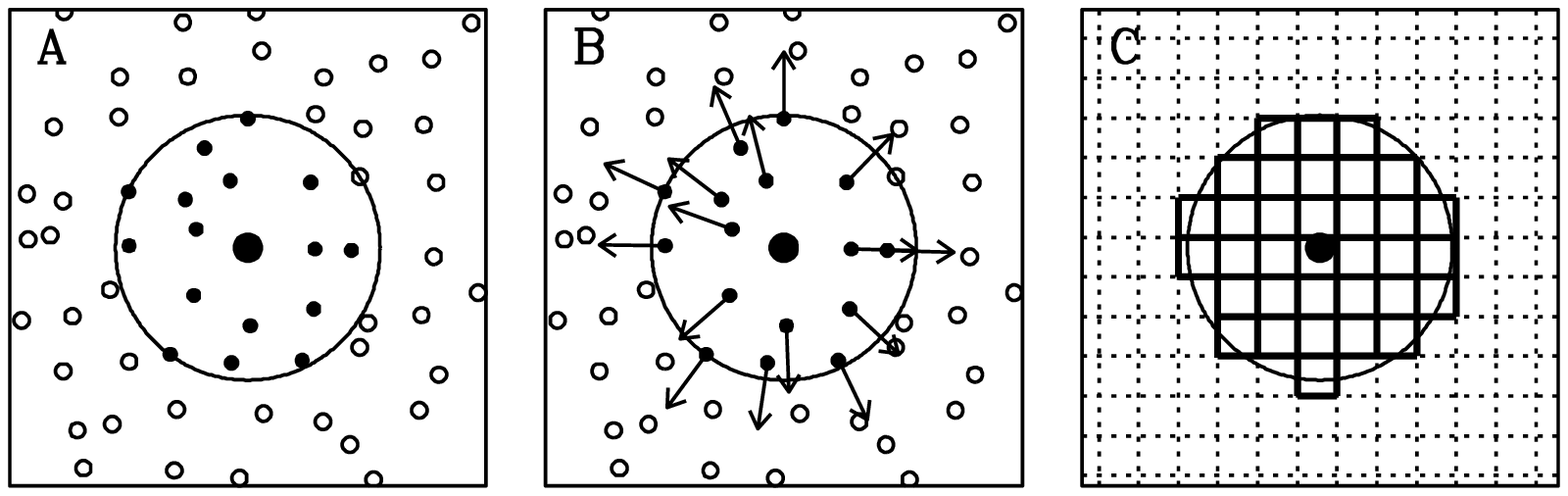}
\end{center}
\caption{\label{feedback}Prescriptions for implementation of feedback in 
numerical simul\-ations. The large black dot shows the object producing
the outflow, and the circle shows the neighboring region where feedback
and metal-enrichment takes place.
A) Thermal feedback in particle-based simulations. Thermal energy is
deposited on neighboring particles (shown in black).
B) Kinetic feedback in particle-based simulations. Neighboring particles
are given an outward kick.
C) Thermal and kinetic feedback in grid-based simulations. Energy is
added to neighboring cells (shown by thick lines).
}
\end{figure}

In the subgrid approach, the algorithm
identifies sites of star and/or galaxy formation. The algorithm
then creates an object that represents structures at unresolved scales.
In Lagrangian, particle-based simulations, these objects can
be {\it star particles}, that represent a population of 
stars\cite{katz92,od06},
or, at larger scales, they can be {\it Galaxy Objects} (GALOBs), that represent
groups of galaxies\cite{jbm11}. In Eulerian, grid-based methods, the algorithm
creates {\it Galaxy Constructs} (GALCONs), subgrids 
that represent individual galaxies\cite{arn10}. 

Once these objects are created, they will affect the surrounding IGM,
in the form of feedback and chemical enrichment. 
In Lagrangian algorithms,
the algorithm identifies the gas particles located within a certain distance of
each object. The metallicity of these particles is then increased, to reflect 
the enrichment caused by the outflow. There are two basic approaches used
to handle feedback: thermal and kinematic. In the first case, the temperature 
of the nearby particles is increased\cite{theunsetal02}. 
In the second case, the nearby particles
are given an outward ``kick'' \cite{od06,std01,sh03}.
This is illustrated in the first two panels
of Figure~\ref{feedback}. Notice that these two approaches often produce
similar results. Increasing the temperature of the nearby particles results in
a pressure gradient that will accelerate particles outward.

In Eulerian algorithms, we are dealing with cells and not particles.
The algorithm identifies the cells located within a
certain distance of each object. Chemical enrichment and feedback is
then handled as in the Lagrangian algorithms. The metallicity in each
nearby cell is increased, and either the temperature or the velocity in
each cell is modified to account for feedback\cite{cno05,kollmeieretal06}. 
This is illustrated in
the last panel of Figure~\ref{feedback}.

In all cases, the algorithm directly calculates the propagation of the 
outflow. Only the source of the outflow (stars or galaxies),
their formation, energy production, and metal production, is treated using
subgrid physics.

\subsection{The Freeze-out Approximation}

There is a third, radically different approach for including galactic 
outflows in cosmological simulations\cite{aguirreetal01}.
Galaxies are identified in
the output of an SPH simulation. Then, an algorithm calculates 
the propagation of outflows
from these galaxies in $N$ different directions, by estimating
the resistance encountered by the outflow in each of these directions. 
We refer to this method as the {\it freeze-out} approximation, because
the dynamics of the IGM is not influenced by the outflow. Instead, the
outflow propagates into a ``frozen'' IGM. 
The main limitation of this approach is that, since it uses the output 
of an SPH simulation and introduces the outflows {\it a posteriori}, 
the feedback effect of these outflows cannot be simulated 
(i.e., outflows do not influence the formation of other galaxies). 
The main advantage is that several outflow models can be considered,
and their parameters can be varied, without
having to redo the full cosmological simulation.

\section{Summary}
Galactic outflows play a critical role in the formation of galaxies and
the evolution of the IGM.
Outflows expel interstellar gas into the surrounding IGM,
reducing the mass of the ISM and potentially quenching star formation.
The energy, momentum, and metal-enriched gas deposited into the IGM 
by outflow affects its evolution and the subsequent formation of galaxies,
either by stripping protogalaxies of their gaseous content or by modifying
the cooling rate of protogalactic gas.

Analytical models and numerical simulations of outflows 
produced by isolated galaxies are useful but have important
limitations. They do not describe the evolution of the IGM and the
subsequent formation of new galaxies,
they do not account for the effect of surrounding structures on the 
propagation of outflows, they often ignore the replenishment of ISM
by accretion of gas from nearby structures, and finally
they do not describe the global effect of an entire galaxy population 
on the evolution of the IGM.

Cosmological simulations are necessary to address these issues, but
purely numerical simulations lack the necessary resolution to properly resolve
cosmological and galactic scales simultaneously.
The solution consists of combining cosmological simulations with 
a subgrid treatment of physical processes at galactic scales.
This can be done using either an analytical outflow model
or a subgrid numerical treatment of feedback and metal enrichment 
by outflows. Such simulations can describe the propagation of outflows and
their impact on the evolution of the IGM inside an entire cosmological
volume containing a representative part of the universe.

\end{document}